# CONTROLLED MOBILITY FOR C-V2X ROAD SAFETY RECEPTION OPTIMIZATION

Jingxuan MEN[1], Yun Hou[2]

[1,2] Department of Computing, The Hang Seng University of Hong Kong, HK

NOTE: Corresponding author: Yun Hou (aileenhou@hsu.edu.hk)

***Abstract*** *– The use case of C-V2X for road safety requires real-time network connection and information exchanging between vehicles. In order to improve the reliability and safety of the system, intelligent networked vehicles need to move cooperatively to achieve network optimization. In this paper, we use the C-V2X sidelink mode 4 abstraction and the regression results of C-V2X network level simulation to formulate the optimization of packet reception rate (PRR) with fairness in the road safety scenario. Under the optimization framework, we design a controlled mobility algorithm for the transmission node to adaptively adjust its position to maximize the aggregated PRR using only one-hop information. Simulation result shows that the algorithm converges and improve the aggregated PRR and fairness for C-V2X mode broadcast messages.*

## 1. INTRODUCTION

In the 5G and future 6G era, everything will be connected through internet. However, communication between vehicles is more challenging than general mobile communications due to fast mobility and high safety requirements. Vehicle-to-Everything (V2X) communication is designed to expand vehicles' sensing ranges beyond the one-hop sensing system provided by on-car camera, radar or lidar. V2X can be regarded as a solution of wireless sensor systems, which allows vehicles to share information with each other through communication channels. It can detect hidden threats, expand the sensing range of autonomous driving, and predict what will happen next to further improve the safety, efficiency, and comfort of autonomous driving. Therefore, V2X is considered to be the key to improve the safety of autonomous driving. When the intelligent connected vehicle communicates with another vehicle, we call it the communication between vehicles (V2V), while when the intelligent connected vehicle communicates with roadside units, traffic lights, road signs, etc., we call it the communication between vehicles and infrastructures (V2I).

In the early days, Dedicated Short Range Communication (DSRC) and Cellular-V2X are two main standards to implement vehicular communications. However, compared with DSRC, C-V2X can support longer communication distance, better performance on NLOS conditions, lower packet lost rate, and higher data rate [1-2]. Due to the limitations of DSRC performance, C-V2X became the a more up-to-date standard for V2X and currently recognized and deployed across the world. C-V2X defines two communication modes for vehicle applications: Base station coordinated Mode 3 and Autonomous Mode 4. In Mode 3, base stations served as the control center to provide scheduling control message for vehicles within the coverage of cellular network. After receiving the control message, the vehicles exchange data information with others through PC5 interface according to the instructions from base station. In Mode 4, the vehicles can operate without a cellular network coverage. Carrier sensing and the Semi-Persistent Scheduling (SPS) procedures are adopted for transmission to avoid potential packet collision without the assistance of base stations. Road safety is one of the major use cases supported by C-V2X, and broadcast is used to cover as many as possible vehicles in the neighborhood.

This paper mainly focuses on the C-V2X road safety use cases, where the status update and warning messages are periodically broadcast. In this case, the per-node packet reception rate (PRR), which is defined as the ratio between the number of neighboring vehicles with successful reception and the total number of neighboring vehicles, is the main objective to maximize for a C-V2X network. In this paper, we study the influence of vehicles geo-location on PRR, and propose an algorithm to optimize the PRR with fairness among neighbors accounted by controlling the mobility of transmitter, i.e., how the transmitter moves.

The rest of this paper is organized as follows. Section 2 introduces the challenges and related work on the controlled mobility for the PRR optimization. Section 3 and Section 4 describe the system settings and the optimization problem formulation, respectively. The proposed algorithm is derived in Section 5. And performance of the algorithm is evaluated in Section 6 with numerical results. Finally, the Section 7 concludes this paper.

## 2. CHALLENGES AND RELATED WORKS

Road safety use cases of C-V2X, where the vehicles send out status-update and warning messages, require the messages to be disseminated to as many neighboring vehicles as possible. The connectivity and coverage issue has been well studied in the literature, while one important degree of freedom has lightly attended. That is the vehicles' mobility. Mobility indeed is an active impacting factor in the connectivity and coverage for V2X networks. It defines the network topology. Moreover, if vehicles could move wisely in time, the network coverage, in terms of packet reception rates (PRR), could be further improved.

The challenges of research work can be summarized as follows.
- In the previous studies, optimization for C-V2X does not include the movement at the network layer.
- The traditional SINR based link model only parameterize distance. However, finding the optimal distances does not necessarily produce an optimal solution in position.
- Most of the optimization needs to know SINR feedback, but the broadcast has no SINR feedback. Therefore, a predictive model for link performance based on the vehicle's position is needed.

The performance of C-V2X mode 4 has been a research focus in the past few years. The relationship of the PRR and the distance between the transmitter and the receiver is studied in [3], it is pointed out that the PRR is positively correlated with the distance between the users. The transmission errors of mode 4, such as those caused by half-duplex propagation and packet conflict, are comprehensively studied in [4]. The author proposes that the occurrence of conflict depends on the performance of link level and the distance between transmitter and receiver. Since media access control (MAC) parameters, such as the probability of retaining wireless resources and resource counters, will affect the duration of wireless resource conflicts. A new algorithm is designed to optimally configure these parameters according to the current vehicle density in [5]. However, C-V2X specific procedures such as blind-detection of PSCCH without knowing the cyclic shift used by the other nodes, autonomous SPS scheduling and carrier sensing to decide the transmission resource, and the non-feedback mechanism all changes the link level sidelink performance. These studies were not carried out based on realistic sidelink performance models of C-V2X mode 4 and this makes it difficult for the results to reflect the true network level performance.

Traditionally, wireless multi-hop networks are considered to be fixed and easy to manage and control. However, in view of the development of Internet of vehicles and UAV, mobile multi-hop network began to receive research attention [6-8]. The impact of node mobility on wireless multi-hop networks resource allocation is in fact two-sided. On one hand, stationary nodes give us convenience in network management and control. For example, nodes' position information does not need to be updated frequently, and there is no fast channel fading caused by relative movement between nodes. However, on the other hand, stationary nodes constrain the exploitable potential of the wireless network. For example, in terms of network capacity, Tse et al. [9-10] pointed out that the random movement of nodes in a wireless network (even if it is only a random movement in a one-dimensional direction) is able to improve the theoretical network capacity from $\sqrt{N}$ to approximately 1, where N is the number of nodes in the network. It is foreseeable that if nodes perform 'wisely' movement in a controlled manner, the capacity of the wireless network should have more room to grow.

As for the problem of controlled movement, the research in [11] studied how the robot can move to the required position autonomously for sensing and data acquisition. The initial and possibly disconnected networks are self-organized via an autonomous movement to form a bi-connected network. In order to use only one-hop information for movement control, achieve maximize network coverage and minimize the total moving distance, Hai proposes a movement control algorithm which simulates the attractive force and repulsive force in nature, so that each robot only needs to follow the synthetic virtual force to move. The paper finds a way for each robot to control its own movement distributedly. However, it assumes a simplified link model where the reception rate of a packet is purely defined by the distance, without considering the concurrent interference in a multi-hop network.

Some studies use machine learning to predict and adjust vehicle trajectory more accurately, and give the optimal solution of vehicle trajectory adjustment. With the help of reinforcement learning, the study in [12] realizes the preliminary deployment and trajectory optimization technology of stable communication between train and multiple UAVs under UAV energy constraints. Support vector machine is also used for optimal initial deployment according to the maximum UAV communication distance data of train speed and UAV energy. However, this research is mainly to provide 5G based VR / AR experience for passengers on the high-speed trains, without considering road safety where the packet reception rate is the true objective to guarantee URLLC communication. In order to reduce the collision probability between vehicles and vulnerable road users (VURs), consider using the mobile phone position of VURs, a new vehicle service based on regression algorithm is proposed, which uniquely uses Cartesian coordinates to predict the trajectory of vehicles and VURs [13]. The above two methods do not consider the use of C-V2X in vehicular network.

In order to acquire specific link performance in C-V2X network, it is required to set up a physical layer abstraction model to identify the co-channel interference. Co-channel interference comes from all other concurrent transmissions that physically use the same frequency and time resource as the link under consideration. However, previous works made assumption that all the concurrent transmissions are interfering sources, regardless of whether they are transmitting on the same frequency sub-channels or not [14-15]. This assumption works well for Wi-Fi based system but fails to model the interference behavior of C-V2X systems. This is because generic Wi-Fi at 2.4 GHz only has three non-overlapping channels and so it is hard to allocate resources in terms of frequency. All transmitting nodes are regarded as co-channel interferers to others in general. On the contrary, in C- V2X systems, resources are divided in Resource Blocks (RBs) frequency and time in the OFMDA structure. Nodes may transmit at the same time but using different 'resource blocks' in the frequency domain. In this case, concurrent transmissions are not necessarily interfering with each other. The physical layer performance of several vehicle to vehicle (V2V) communication technologies is evaluated and compared [16-18]. However, there is no research work on realistic link modeling of C-V2X to assist the self-optimization of vehicular networks.

In this paper, we formulate the optimization problem of packet reception rate maximization for the road safety scenarios using a C-V2X sidelink mode 4 abstraction and regression results from a C-V2X network-level simulation. Under the optimization framework, we devise a controlled mobility algorithm for transmission nodes to adaptively adjust its position to maximize the aggregated PRR and Utility Gain using one-hop information only.

## 3. SYSTEM SETTING

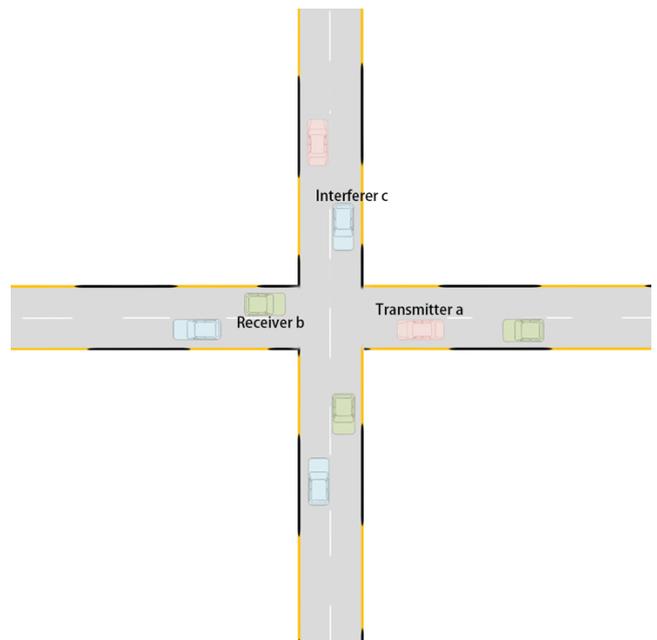

**Fig. 1** – C-V2X communication system model.

As shown in Fig. 1, we consider a square Region of Interest (RoI) consisting 4x4 two-way roads. $N$ vehicles are dropped on the roads and move at a random speed. In the road model, the relative distance $d_{a,b}$ and relative velocity $v_{a,b}$ between transmitter $a$ at position $(x_a, y_a)$, with velocity $\vec{v_a}$, and receiver $b$ at position $(x_b, y_b)$, with velocity $\vec{v_b}$ could be calculated according to:

$$d_{a,b} = \sqrt{(x_a - x_b)^2 - (y_a - y_b)^2} \quad (1)$$

$$v_{a,b} = \sqrt{(\vec{v_a} - \vec{v_b})^2} \quad (2)$$

Next, the data obtained from the road model could be used to the network model. It is assumed that the transmission power of transmitter $a$ is $s_a$, the carrier frequency of the signal is $c$, and the noise power $n_p$ is considered as Gaussian White Noise. Therefore, pathloss from vehicle $a$ to $b$ is $\rho_{a,b}$ and the doppler between them is $D_{a,b}$:

$$\rho_{a,b} = K \cdot \left[\frac{d_0}{d_{a,b}}\right]^\omega \quad (3)$$

$$D_{a,b} = \frac{2 \cdot c \cdot v_{a,b}}{3 \times 10^8} \quad (4)$$

The pathloss model indicates that the pathloss gain for link from $a$ to $b$ is proportional to the order $\omega$ of $1/d_{a,b}$, where $\omega$ equals to 2 in the free space scenarios. Then the resource selection process of SPS could be simulated. First, calculate the measured Reference Signal Received Power (measured RSRP), threshold Reference Signal Received Power (threshold RSRP) and Received Signal Strength Indicator (RSSI) of the resources, and then make a list. The autonomous scheduling with three steps according to the SPS mechanism is as follows.

- If an SCI is received on the time slot, or the corresponding time slot is reserved by the previous SCI, and the measured RSRP is larger than the threshold RSRP, the resource is excluded. The remaining candidate resources should be 20% of the total resources in the resource selection window. Otherwise, increase the threshold RSRP by 3dB until the conditions are met.
- Sort the candidate resources according to their RSSI, and select the best three resources.
- Randomly select one of the three resources.

After a packet is transmitted using the chosen resource, the received signal power $r_{a,b}$ of link $l$ is computed as:

$$r_{a,b} = s_a \cdot \rho_{a,b} \cdot ch_{a,b}^2 \quad (5)$$

The interference power $I_{c,b}$ from one of interferers $c$ to receiver $b$ is calculated as:

$$I_{c,b} = s_c \cdot \rho_{c,b} \cdot ch_{c,b}^2 \quad (6)$$

Where $s_c$ is the transmission power of $c$, $\rho_{c,b}$ is the pathloss from vehicle $c$ to $b$. And the total interference affecting the signal transmitted on link (a, b) is:

$$I_{a,b} = \sum_{c \in Tx} I_{c,b} \quad (7)$$

In (7), the summation suggests that the total interference affecting the signal transmitted on link (a, b) is coming from the set of all nodes that are concurrently transmitting at the same time, i.e., the set denoted by **Tx**. The channel fluctuation in time and frequency, i.e., compliant to winner model, $ch_{a,b}$ is the mean channel coefficient between $a$ and $b$, and $ch_{c,b}$ is the mean channel coefficient between $c$ and $b$. The mean is applied across all REs in reference symbols of that packet. Then the SINR of link $a$ to $b$ could be calculated as:

$$SINR_{a,b} = \frac{r_{a,b}}{I_{a,b} + n_p} \quad (8)$$

In order to obtain the link success probability $P_{a,b}$ of each vehicle communication link, considering the randomness of the vehicle network channel, we perform per-link bit level simulations on discrete SINR and velocity points, and obtain curves of link success probability as function of the Doppler and the SINR by linear interpolation. The interpolated curves form the C-V2X physical abstraction and is used to obtain a the success probability for a transmission from vehicle $a$ to vehicle $b$ using the Doppler and SINR calculated from (4) and (8), respectively. For a transmitter, the transmission power $s$ values returned by each transmission link could be added to obtain the per-node PRR. As our current study focuses on low-modest speed scenario with high vehicle density, i.e., urban scenarios, the change of doppler has little effect on link reception success probability. Therefore, the velocity is not included in the physical abstraction model in the following of the discussion. The link success probability is purely defined by the SINR of the link, whereas the channel fluctuation is still affected by the Doppler.

## 4. PROBLEM FORMULATION

In order to maximize the reliability for the road safety in urban scenarios, a broadcast packet is desired to be received by neighboring vehicles as much as possible. While, when one vehicle is transmitting, the rest of N-1 vehicles in the RoI may decide to transmit or receive in that subframe. Due to the half-duplexing operation mode, only the vehicles in a receiving mode are considered to be the targeted receivers. Thus, assume there is a utility associated to each target receiver, i.e., to represent a level of satisfaction that a target vehicle can obtain from successfully receiving a packet. Then, at the transmitter, its objective will be to move to a better position such that the aggregated utility of its target receivers is maximized. In this way, we can formulate node $i$'s aggregated utility as

follows:

$$U_i = \sum_{j \in \mathbf{R_i}} P_{i,j}, \quad (9)$$

Where $P_{i,j}$ stands for the success probability of the reception on link from node *i* to node *j* and $\mathbf{R_i}$ is the set of all receiving nodes in the neighborhood of *i*.

The exact form of function $P_{i,j}(.)$ is yet to be found via a Machine Learning approach in section 5. However, we will not lose any generality to assume that $P_{i,j}$ is a monotonic function of the link $SINR_{i,j}$ defined in (8). Ignoring the influence of Gaussian noise, the SINR in (8) is directly proportional to the signal power of the receiver and inversely proportional to the sum of the interference caused by the interferers. Thus, we can derive that the $P_{i,j}$ has a negative relation to $d_{i,j}$, where $d_{i,j}$ is the distance between the transmitting node *i* and the receiving node *j*. Considering the pathloss definition in (3), we have:

$$\frac{\partial P_{i,j}}{\partial d_{i,j}} \propto \frac{1}{d_{i,j}}. \quad (10)$$

On the contrast, $P_r$ should have a positive relation to the distance between the receiver *j* and its interfering nodes, i.e., transmitting nodes that are close enough (lies in a distance smaller than a pre-defined threshold $th_I$ to the receiver *j*). Although for one receiver node *j*, there could be more than one interferers, it could be meaningful to focus on the closest interferer of the receiver *j* to simplify the problem model. Let us denote the distance between the receiver *j* and its closest interferer as $l_j$. Then, similar to (10), we can have:

$$\frac{\partial P_{i,j}}{\partial l_j} \propto l_j. \quad (11)$$

Let us refer $d_{i,j}$ and $l_j$ as signal distance and main interference distance hereafter. Then, the two gradient (10-11) depicts the impact of the two distances on the link reception rate. That is, driving by the gradient inversely proportional to $d_r$ in (10), the transmitter benefits from reducing the distance $d_r$ from all the receivers. Moreover, before the transmitter moves, if a target vehicle is closer to a receiver, it will be more beneficial to move even closer to that receiver so that the link reception rate $P_{i,j}$ or $U_i$ will be maximized. In this way, the node gains little by moving towards the receivers at the far side, and in a distributed manner the node will not have the motivation to move towards the far nodes to improve their utility. Thus, this is a greedy utility function without any fairness consideration between target vehicles.

As a result, a fair utility function should be considered to shrink the gain in distance when the $P_{i,j}$ is already high. We propose to apply log for the $P_{i,j}$ in the function (9) and get:

$$U_i = \sum_{j \in \mathbf{R_i}} \log(P_{i,j}) \quad (12)$$

With the log function, the gradient with respect to $d_{i,j}$ becomes:

$$\frac{\partial U_i}{\partial d_{i,j}} = \frac{1}{P_{i,j}} \cdot \frac{\partial P_{i,j}}{\partial d_{i,j}} \quad (13)$$

It means that the increase in terms of $U_i$ in (12) that can be contributed from an increase in $d_{i,j}$ is inversely proportional to the current $d_{i,j}$ as well as its current $P_{i,j}$. The sign of (13) may be the same among different receiving vehicles, which means by getting closer to any target receiver, it is always bring up the total utility. While the amplitude of the partial derivatives varies from receiver to receiver. That means, by moving closer to one target receiver by a unit distance, the return in aggregated utility at the transmission node for the whole broadcast packet, is different from what brought by moving closer to another target receiver. In this way, the transmitter can decide which target receiver is the most rewarding target to move closer to and hence a gradient-based updating method may be derived from there to move the transmitter bit by bit iteratively to its optimal position.

Thus, the optimization is finally formulated as:

$$\max_{d_{i,j}, j \in \mathbf{R_i}} U_i = \sum_{j \in \mathbf{R_i}} \log P_{i,j}(d_{i,j}, l_j) \quad (14)$$

and an updating algorithm to solve this optimization problem by adjusting $d_{i,j}$ is derived in the following Section.

## 5. THE PROPOSED ALGORITHM

To obtain the real performance of each communication link, we need to use V2X-specific and accessible attributes to represent link performance. With the discussion above, the per node PRR should be predictable based on the two distances, e.g., the signal distance and the main interference distance. we propose a **two-distance model** to predict $P_{i,j}$ for a receiver as follows:

$$P_{i,j}(d_{i,j}, l_j) = \alpha_j \log d_{i,j} + \beta_j \log l_j + \gamma_j \quad (15)$$

where $\alpha_j$ is the coefficient associated to signal distance $d_{i,j}$, $\beta_j$ is the coefficient associated to main interference distance of a receiver *j*, i.e., $l_j$, $\gamma_j$ is the intercept from the regression.

The values of $\alpha_j$, $\beta_j$ and $\gamma_j$ are receiver-specific coefficients as their values depend on how many interferers are surrounding $j$ as well as the strengths of the interferences. It is because that the simplified formula of $P_{i,j}$ considers the main interference distance only. The accuracy of using the main interference to represent the whole interference in the SINR term in (7-8) varies when the composition of interference at $j$ varies. Thus, we propose to assume the coefficients $\alpha_j$, $\beta_j$ and $\gamma_j$ are dependent on the Number of Surrounding interfering Vehicles (NSV). The value of $\alpha_j$, $\beta_j$ and $\gamma_j$ will be found via Regress in Section 6 for different number of NSVs.

With (15), the three-dimensional relation curve of signal distance, main interferer distance and success probability could be plotted in Fig. 2. It can be seen from the figure that $P_{i,j}$ is a monotonically growing function with signal distance. Therefore, for each vehicle, shortening the signal distance will result in better link success probability. However, the slope will be vanished, which means that when the distance is larger, the gain by making it shorter is not that much. While, when the distance is small (close to zero), it will be very attractive to make it even shorter.

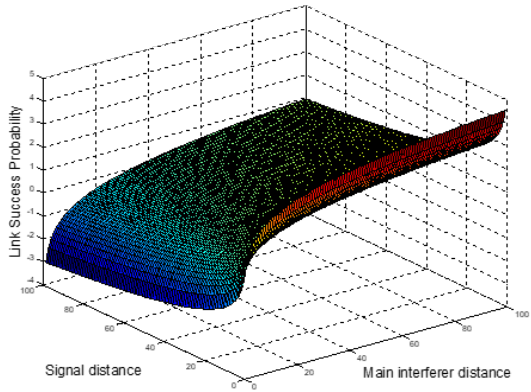

**Fig. 2** – Three-dimensional relation curve of Signal distance, Interferer distance and Success Probability

By taking partial derivative on both sides, we have the gradient of the utility with respective to $d_{i,j}$ as:

$$G_{i,j} = \frac{\partial U_i}{\partial d_{i,j}} = \alpha_j \cdot \frac{1}{d_{i,j}} \cdot \frac{1}{P_{i,j}} \quad (16)$$

The gradient $G_{i,j}$ indicates that the transmitter moves in a fair manner, i.e., if the $P_{i,j}$ of one link for a receiver is high, the gain from moving closer will be relatively small.

For example, $UE_0$ is the transmitter in the Fig. 3, $UE_1$, $UE_2$ and $UE_3$ are three receivers of the current V2X packet, they are traveling in three directions. $d_{0,1}$, $d_{0,2}$, $d_{0,3}$ ($d_{0,3} > d_{0,1} > d_{0,2}$) are the distance between transmitter $UE_0$ and receivers. The $Interferer_4$, $Interferer_5$ and $Interferer_6$ are three interferers using the same resource with transmitter $UE_0$. $l_1$, $l_2$ and $l_3$ ($l_2 > l_1 > l_3$) are the distance between interferers and receivers. $UE_2$ has a good packet receiving probability because the signal distance is small while the interference distance is big. Similarly, we can find that $UE_1$ has a modest receiving probability while $UE_3$ has a lowest receiving probability. In order to optimize the PRR, the $d_{0,1}$, $d_{0,2}$ and $d_{0,3}$ need to be adjusted by moving the transmitter. The higher the return it is estimated to have from the moving closer to a receiver $j$, the more the transmitter should move towards the receiver $j$.

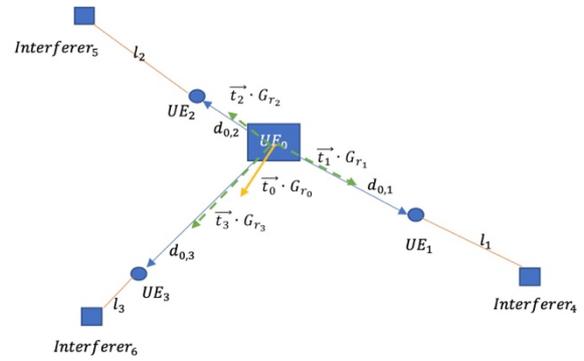

**Fig. 3** – The position of transmitter, receivers and interferers

An iterative updating method for position, called **Gradient-Based-Mobility-Updating** (GBMU) Algorithm, is proposed to move a broadcast transmitter $i$ and it consists of five steps:

- Step 1: Transmitter $i$ obtain gradient $G_{i,j}(k)$ of the current iteration $k$ for all target receiver $j$ by:

$$G_{i,j}(k) = \alpha_j \cdot \frac{1}{d_{i,j}(k)} \cdot \frac{1}{P_{i,j}(k)}$$

  where $\alpha_j$ is the coefficient gained from the regression step, depending on receiver $j$'s situation, i.e., how many neighboring interfering vehicles is in $j$'s range. The reason is, the regression resultant coefficient varies when the number of selected vehicles is different, referring Table 2 in Section 6.

- Step 2: Transmitter calculate the unit vector to indicate the moving direction for all receiver $j$ by:

$$\overrightarrow{t_j(k)} = [Pos_i(k) - Pos_j(k)]/d_{i,j}(k)$$

  Where $Pos_i$ is the position of transmitter, and $Pos_j$ is the position of receiver. Both $Pos_i$ and $Pos_i$ are vectors containing the x- and y-coordinates.

- Step 3: obtain the wanted position of transmitter with respect to $j$ by moving a distance along the calculated direction, for all $j$. The distance is proportional to the gradient of $U_i$ with respect to the current $d_{i,j}$.

$$Pos_{i,j}(k+1) = Pos_i(k) + \delta \cdot \overrightarrow{t_j(k)} \cdot G_{i,j}(k)$$

- Step 4: assign the updated position of $i$ as the centroid of $Pos_{i,j}(k+1)$ across all $j$'s
$$Pos_i(k+1) = \underset{j}{\text{mean}}[Pos_{i,j}(k+1)]$$

- Step 5: if $Pos_i(k+1) - Pos_i(k) < Th_{Pos}$, terminate the algorithm and move to the optimal position, otherwise go back to step 1.

In the above steps, $\delta$ is the constant step size and $Th_{Pos}$ is the a small threshold for algorithm to stop. These two constants are empirical values gained from simulation in Section 6. In the above example illustrated in the Fig. 3, $\overrightarrow{t_1}, \overrightarrow{t_2}, \overrightarrow{t_3}$ are the moving directions from $UE_0$ to receivers $UE_1, UE_2$ and $UE_3$. $G_{r_1}, G_{r_2}, G_{r_3}$ are the gradient for receivers obtained by $UE_0$ in step 1 and 2. Therefore, $\overrightarrow{t_1} \cdot G_{r_1}, \overrightarrow{t_2} \cdot G_{r_2}, \overrightarrow{t_3} \cdot G_{r_3}$ are the moving distance towards $UE_0$ to $UE_1, UE_2$ and $UE_3$, if $\delta = 1$ in step 3. The final movement in step 4 for this iteration will be the one illustrated with the orange arrow annotated with directio $\overrightarrow{t_0}$ and gradient $G_{r_0}$. It is the centroid of the three desired positions given by the green arrwos, taking the tradeoff of PRR vs fairness.

### *The availability of the parameters in GBMU:*

Throughout step 1-5, our GBMU algorithm requires the transmitter to know for each j, a) the position of j, b) the position of the main interferer of j, c) number of surrounding interfering vehicles (NSV) so that the associated coefficient values to that NSV could be known. For a) and b), the position information is a well-exchanged information in C-V2X. For the knowledge on b) and c), it could be inferred from previously received SCI packets given that the SPS scheduling used in C-V2X offer high periodicity in transmission patterns.

The **merit of the proposed GBMU scheme** is that the prediction of the PRR and hence the utility can be obtained based on a simple Linear Model and locally and readily obtainable information, i.e., the signal distance and the main interference distance, which are well available for V2X communications as the geo-location information are always exchanged by the vehicles. The Linear Regression model in (15) and the GBMU updating method release the communication entities, i.e., UEs where computations power and knowledge of the network is very limited in C-V2X mode 4, from complex and expensive signaling for SINR feedbacks.

## 6. SIMULATION RESULTS AND DISCUSSION

### 6.1 Simulation setting

A 2-stage simulation has been conducted to verify the proposed algorithm as depicted in Fig. 4.

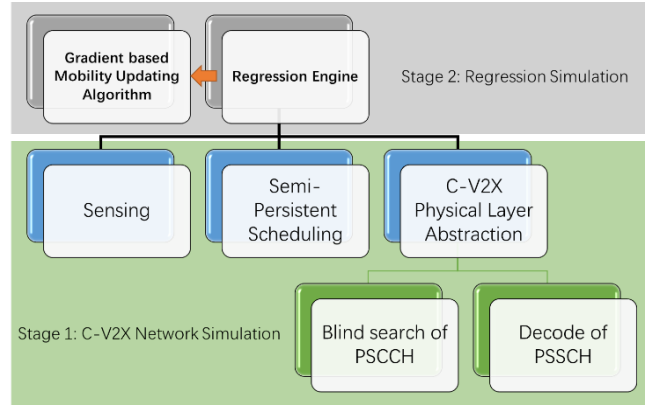

**Fig. 4** – Simulation hierarchy

The stage 1 is a network level simulation with the following parameters shown in table 1 to mimic a network of 21 vehicles moving with random speed on a RoI of 450m * 450m as described in Section 3. In the network simulation, in a snapshot subframe, there may be multiple vehicles transmitting, driven by a packet generator with uniform packet arrival rate and sidelink sensing and autonomous scheduling fully compliant to C-V2X standard [19]. With all the concurrent transmission accounted as interference, the SINR and relative velocity of a link can be computed and used in to obtain the packet success rate. The relation from SINR to successful reception rate $P_{i,j}$ is provided by a sidelink abstraction simulation with link-level Matlab simulations conforming to [20]. The key parameters used in our Matlab simulations are shown in Table 1.

**Table 1** – Simulation parameters

| Parameters | Value |
| --- | --- |
| Number of vehicles | 21 |
| Transmission power $s$ | 26 dBm |
| carrier frequency $c$ | 2GHz |
| Noise power $n_p$ | -112.45 dBm |
| Received noise figure | 9 dBm |

| Number of total subframe | 10240 |
| --- | --- |
| Number of total channels | 2 |
| Channel | Winner model |
| Resource selection window | 8 subframes & 2 channels |
| Neighboring interferer distance threshold | 200m |

The stage-1 simulator generates the data needed for stage-2 to conduct regression algorithms to get the two-distance model of the link reception probability in (15). The data attributes are composed of the signal distance, the main interference distance as well as the number of surrounding vehicles that are transmitting simultaneously (NSV). With the obtained regression model in (15), another simulation is performed to verify the performance of the proposed GBMU algorithm with both illustrative scenario (i.e., a 7-node network) as well as the simulation scenario used in stage-1 (i.e., a 21-node network).

## 6.2 Regression performance analysis

We use linear regression to obtain different prediction functions according to the number of interferers around the vehicle in Python. In this linear regression model, R-square is considered as the evaluation standard of fitting. R-square refers to the goodness of fit, which is the fitting degree of the regression line to the observed value. That is, the closer the value of R-square is to 1, the better the fitting degree of the regression line to the observed value is. A R-square greater than 0.4 suggests a meaningful regression.

Our linear regression is carried out with respect to various NSV values (which is the number of interferers within an awareness range of 200 meters nearby). The resultant coefficients associated with the attribute is shown in **Table 2**, with the regression performance metrics shown in table 3.

**Table 2 – NSV-specific Coefficients**

| NSV | 0 | 1 | 2 | 3 | 4 | 5 | 6 |
| --- | --- | --- | --- | --- | --- | --- | --- |
| $\alpha_j$ | -0.83 | -0.71 | -0.61 | -0.67 | -0.72 | -0.76 | -0.74 |
| $\beta_j$ | 0.52 | 0.19 | 0.12 | 0.09 | 0.06 | -0.02 | 0.25 |
| $\gamma_j$ | 1.00 | 1.42 | 1.26 | 1.44 | 1.54 | 1.73 | 1.25 |

In this study, R-square is used as the main evaluation metric. In Fig. 4, it can be observed that with the increase in the number of interference sources within 200m, the R-square value of linear regression gradually increases and reaches the maximum R-square value of 0.72 when NSV = 5. When the NSV is equal or greater than 6, the value of R-square decreases slightly due to lack of data samples. Considering that the probability of more than 6 transmitters select the same resource at the same time is very small, then the case of more than 6 NSVs are not considered. When NSV = 0, there is no interfering vehicles nearby, random channel changes dominate the results, therefore, the R-square value currently is lowest.

**Table 3 – Regression Statistics with NSV-specific coefficients**

| NSV | 0 | 1 | 2 | 3 | 4 | 5 | 6 |
| --- | --- | --- | --- | --- | --- | --- | --- |
| Instances | 321 | 584 | 598 | 418 | 203 | 49 | 8 |
| R Square | 0.28 | 0.46 | 0.51 | 0.59 | 0.68 | 0.72 | 0.46 |

The high R-square score shows that our proposed algorithm is effective in high vehicle density scenarios where the resource pool is highly utilized and more than one neighboring vehicle needs to share the same resource. It is worth noting that this kind of scenario may map to congested roads in reality where road safety messages have a pressing need to be disseminated to more neighbors in a fair manner.

## 6.3 The GBMU algorithm simulation results

As can be seen in Fig. 5, the location of transmitter (denoted by "UE0" in the figures) is updated iteratively according to the algorithm with step size $\delta = 1$. UE0 finally reaches a point in the network and the moving distance per iteration diminishes as the iteration proceeds. In order to verify the convergence and the effectiveness of the algorithm, after each movement of UE0 in iteration $k$, the new position will be used to calculate the path loss $\rho_{i,j}$ and so on the $P_{i,j}(k)$ of each link, and finally the normalized Utility Gain at iteration $k$ could be computed as:

$$gain_i(k) = \frac{U_i(k) - U_i(0)}{|U_i(0)|} \quad (17)$$

In this function, $U_i(k)$ is the new Utility Gain and $U_i(0)$ is the initial Utility Gain. The results obtained are shown in Fig. 6. It can be observed that after 500 times position updates, the algorithm shows convergence and achieves a utility gain of 42.2% in this case.

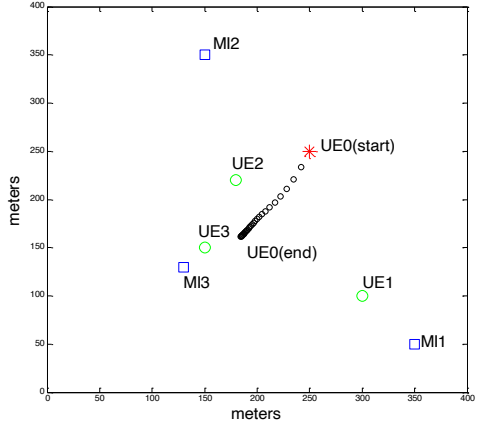

**Fig. 5** – Moving update model

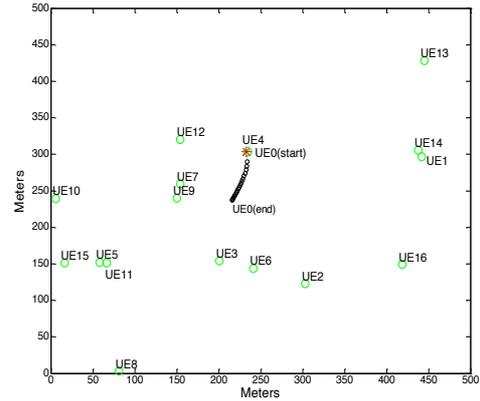

**Fig. 7** – Moving update model with all vehicles

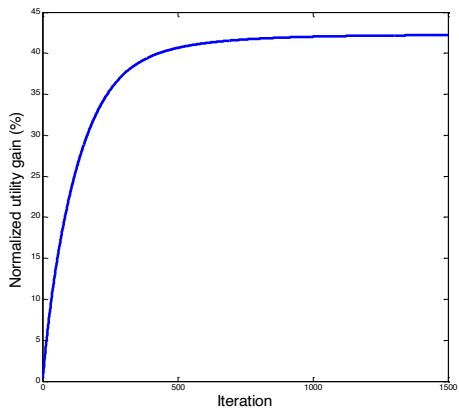

**Fig. 6** – Gain Convergence

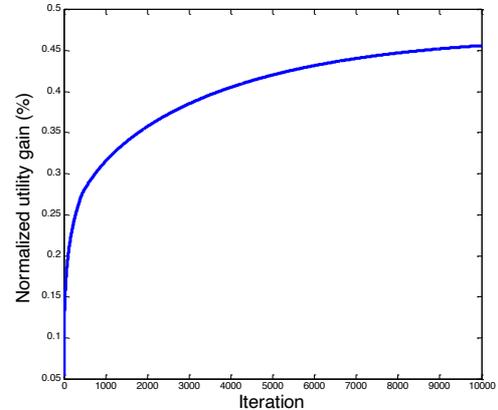

**Fig. 8** – Gain Convergence with all vehicles

Next, we apply the GBMU algorithm to the vehicles simulated in our stage-1 network level simulator, where there are 21 UEs in the network and UE0 is the vehicle whose packet reception rates before and after the moving updates are recorded to visualize the convergence. Fig. 7 shows the iterative position updates with the step size $\delta = 1$ for one the snapshots where UE0 is originally at a non-centered position with respect to all other vehicles. With GBMU algorithm, UE0 can find itself a better position to stay in order to maximize its utility, which takes the success packet reception probability for all its potential receivers and fairness among them into account. The GBMU algorithm takes UE0 finally to reach a better position in this vehicle network and stay there as the algorithm smoothly converges with an utility increase by 45.6% at its convergence shown in Fig. 8. Please note that the final position can be found by the node UE0 via in-node computation to go over the iterations, instead of really moving itself in the network before the convergent position is obtained.

Fig. 9 and Fig. 10 show the converged PRR and utility gain obtained by UE0 for 144 realizations, i.e., 144 initial deployment of UEs in the network. For all realizations, we use step size = 1 and 10,000 iterations. After executing our algorithm, the average PRR increases over all realizations is 5%. It could be seen from Fig.9 that for 134 cases, our GBMU algorithm enhances UE0's utility and only in 10 realizations, the GBMU algorithm leads the UE0 to a worse position. The effectiveness rate of our algorithm is 93% in terms of utility improvement. The overall utility enhancement, averaged across all realizations, is by 15%.

In terms of the PRR of UE0 before and after the position updates provided by GBMU, the average gain across all realization is 5%. There are 40 realizations where the updated UE0's position causes a PRR drop compared to that obtained from UE0's original place. It is because that we use the fairness-oriented Utility to drive the node moving direction and amplitude, the PRR will be sacrificed in some cases, but the obtained utility gain increases indeed.

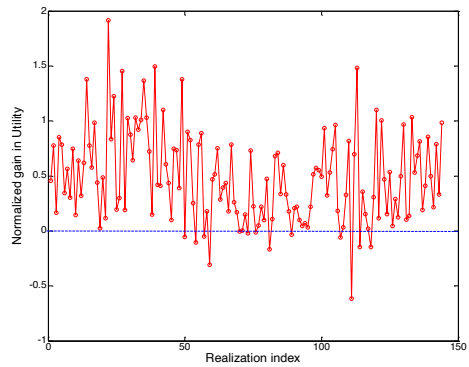

**Fig. 9** – Normalized gain in Utility at convergence across 144 realizations

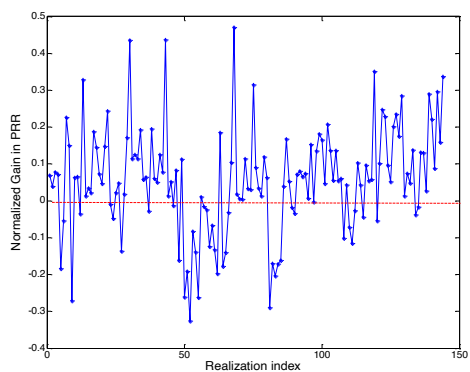

**Fig. 10** – Normalized gain in PRR at convergence across 144 realizations

## 7. CONCLUSION

In this paper, we obtained a two-distance link reception rate model for C-V2X based Linear Regression with results from C-V2X network-level simulation and formulate an optimization problem for per-node PRR maximization with fairness for the broadcasting scenarios. An updating algorithm (GBMU) was devised to solve the optimization problem iteratively, by controlling the mobility of the transmitter. Simulation results show that our two-distance model has high accuracy in scenarios where the C-V2X network has high vehicle density and more concurrent transmissions using the same resource. The proposed GBMU is demonstrated to converge and provide an improvement in per-node utility by 45%.

## AUTHORS

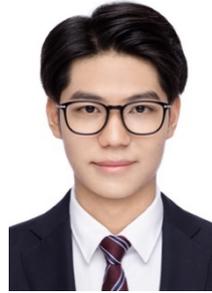

**Jingxuan MEN** received the B.Eng. degree from China University of Geosciences (Beijing), China in 2019 and the M.Sc. from Hong Kong University of Science and Technology, Hong Kong, China in 2021. He is currently working as a Research Assistant at The Hang Seng University of Hong Kong. His research interests are wireless communication and vehicular networking.

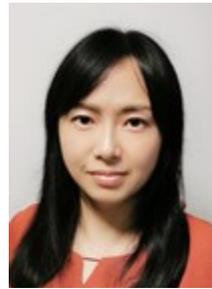

**Yun Hou** received her PhD degree in Electrical and Electronic Engineering from Imperial College London, UK in 2009. She is currently an assistant professor with the Department of Computing, The Hang Seng University of Hong Kong. Before that, she had been conducting applied research in Hong Kong Applied Science and Technology Research Institute (ASTRI) as a Senior Lead Engineer on 3G, 4G and 5G mobile communication systems. She also had working experience in Shenzhen University, IBM and Alcatel-Lucent with research focuses on V2X, Smart City, 5G, Network Optimization and Machine Learning. Her research interests include vehicular communication networks, optimization and machine learning.